\documentstyle[preprint,prl,aps]{revtex}
\begin{document}
\draft
\title{Interaction of Ultrasound with Vortices in Type-II Superconductors}
\author{E.B. Sonin} 
\address{Ioffe Physical Technical 
Institute, 
194021 St. Petersburg, Russia}
\date{\today}
\maketitle

\begin{abstract}
The theory of the ultrasound propagation in the mixed state of type-II
superconductors is suggested which takes into account the Magnus force on
vortices, the anti-Magnus force on ions, and diamagnetism of the mixed
state. The acoustic Faraday effect (rotation of polarization of the
transverse ultrasonic wave propagating along vortices) is shown to be linear
in the Magnus force in any regime of the flux flow for wavelengths used in
the ultrasound experiments now. Therefore, in contrast to previous
predictions, the Faraday effect should be looked for only in clean
superconductors with a large amplitude of the Magnus force. 
\end{abstract}

%\narrowtext

Investigation of ultrasound propagation in the mixed state of a
type-II superconductor has proved to be an effective method of
studying high-T$_c$ superconductors \cite{P} (see also Ref. \cite{SN}
for ultrasound experiment and theory in low-T$_c$ superconductors).
Recently Dominguez {\it et al.} \cite{DB} have attracted attention to
an interesting manifestation of interaction between the ultrasound
and vortices: {\em acoustic Faraday effect}. The effect results from
a force on ions transverse to the ion velocity. Due to this force the
velocity of the transverse sound propagation depends on the sign of
circular polarization and the plane of linear polarization should
rotate when the sound wave propagates along vortices.

However, the system of equations used by Dominguez {\it et al.} \cite{DB}
(see also their later paper \cite{DB1})  is not Galilean-invariant and does
not satisfy  the momentum-conservation law for the whole system
``ions+electrons'' since they missed to include into the ion equation of
motion the so-called anti-Magnus force which is especially important for
dirty superconductors.  In the present work I use dynamic equations which
are free from these deficiencies. The correct theory predicts other
functional dependences on the physical parameters and different conditions
for observation of the Faraday effect. In particular, our analysis does not
confirm the conclusion of  Dominguez {\it et al.} on the strong Faraday
effect without the Magnus force in dirty superconductors at high
temperatures and predicts a many orders of magnitudes weaker effect at low
temperatures. Finally, for the ultrasound wavelengths used in the
experiments the best conditions for observation of the acoustic Faraday
effect are at high temperatures in clean superconductors. The present work
derives the theory of  Refs. \cite{P,SN} as a particular limiting case, but
generalizes the theory on a much larger domain of physical parameters: the
nonzero Magnus force (clean superconductors), low magnetic fields, where the
diamagnetism of the mixed state is essential, wavelengths short compared to
the London penetration depth. 

We adopt the following picture of electron and ion motion induced by the
ultrasound. As usual, the electron liquid consists of two parts: normal and
superfluid. The normal component is effectively clamped to the crystal ions
by viscous forces responsible for the normal resistance: they move together
with the ion velocity $\vec v_i=d\vec u_i/dt$ ($\vec u_i$ is the ion
displacement). In contrast, the superfluid electrons move with the
superfluid velocity $\vec v_s$ which is different from the ion velocity in
general. Such a physical picture holds until the magnetic field is weak
compared to the upper critical magnetic field $H_{c2}$ and the flux-flow
resistance is much less than the normal resistance. Then the normal current
proportional to the velocity of the normal electrons with respect to ions
may be neglected. 

Thus our three-component system (ions, normal and superfluid electrons)
becomes effectively two-component as in the two-fluid model for superfluids:
there is a superfluid with the velocity $\vec v_s$ and the mass density
$m_en_s$ and a heavy normal fluid with the velocity $\vec v_i$ and the mass
density $m_i n+ m_e(n-n_s) \approx m_in$. The charge densities of these two
components are $en_s$ and $-en_s$. Here $e$ is the electron charge, $m_e$
and $m_i$ are the electron and the ion masses, and $n_s$ and $n$ are the
superfluid electron and the total charge number densities respectively. We
can write the equations of motion for the superfluid and the normal fluid
immediately using a close analogy with the two-fluid hydrodynamics for
rotating superfluids (modifications due to presence of the electromagnetic
forces are self-evident) \cite{S}:
\begin{equation}
\frac{\partial\vec v_s}{\partial t} =
{e \over m_e} \left\{\vec E +{1 \over c}[\vec v_L \times \vec
B]\right\}~. 
          \label{elec} \end{equation}     
\begin{eqnarray}
\frac{\partial^2\vec u_i}{\partial t^2} - c_t^2\vec \nabla^2\vec u_i=
-{e \over m_i} {n_s \over n}\left\{\vec E +{1 \over c}[\vec v_i
\times \vec B] \right. \nonumber \\ \left.
+ {1 \over c}[(\vec v_L-\vec v_{sl}) \times \vec B] \right\}~,
        \label{ion} \end{eqnarray}
Here $c_t$ is the ``bare'' sound velocity ignoring interaction with
superfluid electrons, the equilibrium magnetic induction $\vec B$ is
proportional to the vortex density $B/\Phi_0$, $\Phi_0$ is the
magnetic flux quantum, $\vec v_L=d\vec u_L/dt$ is the vortex
velocity, $\vec u_L$ is the vortex displacement, and $\vec v_{sl}$ is
the {\em local} superfluid velocity determined at the points of the
vortex line which differs from the {\em average} superfluid velocity
$\vec v_s$ because of deformations of the vortex lattice. If the
transverse sound wave propagates along vortices (the axis $\hat z$),
then
\begin{equation}
\vec v_{sl}=\vec v_s +
\frac{c}{en_sB} C_{44}^* \left[\hat z \times \frac{\partial^2 \vec
u_L }{\partial z^2} \right]~. 
          \label{bv}  \end{equation} 

The renormalized tilt-modulus $C_{44}^*$ relates only to the vortex
line-tension, without including the elastic energy of the average
magnetic field \cite{S2}. For an isotropic superconductor not very
close to $H_{c1}$ $C_{44}^*=
(\Phi_0B/4\pi \lambda^2)\ln (a/r_c)$ (for an anisotropic case see,
e.g., \cite{KS} and references therein). Here $a \sim \sqrt{\Phi_0
/B}$ and $r_c$ are the intervortex distance and the vortex core
radius.

The electrical field $\vec E$ and the
magnetic field $\vec h$ generated by the ultrasound wave satisfy the
Maxwell equations:
\begin{equation}
\vec \nabla \times \vec E = -{1 \over c} \frac{\partial \vec
h}{\partial t}~,~~~~~{4\pi \over c}\vec j=
\vec \nabla \times \vec h~,
        \label{max} \end{equation}
where $\vec j=en_s (\vec v_s - \vec v_i)$ is the average electric
current. The superfluid component of the current, the supercurrent
$\vec j_s = en_s \vec v_s$, should satisfy the London equation
averaged over the vortex array cell:
\begin{equation}
\frac{4\pi \lambda^2}{c} \vec \nabla \times \vec j_s=\vec b_v -
\vec h~.
       \label{lon} \end{equation}
Here $\lambda=\sqrt{m_ec^2/4\pi e^2n_s}$ is the London penetration
depth and $\vec b_v $ is the a.c. component of the vortex field
\cite{MS} (the vortex induction in \cite{S1}) which coincides with 
the magnetic field $\vec h$ only in an uniform vortex lattice.  For
the transverse sound propagating along vortices $\vec b_v= B \partial
\vec u_L/\partial
z$.

We need also the equation of vortex motion which connects three
velocities $\vec v_i$, $\vec v_{sl}$, and $\vec v_L$:
\begin{eqnarray}
-\eta (\vec v_L - \vec v_i) + \eta '[\hat{z}\times (\vec v_L
- \vec v_i)] \nonumber \\
=\pi \hbar n_s[\hat{z}\times (\vec v_{sl}- \vec v_i)]~.
           \label{vort} \end{eqnarray}
Equation (\ref{vort}) is a self-evident generalization of the
equation of the vortex motion known for a crystal at rest: the
velocities $\vec v_L$ and $\vec v_{sl}$ are replaced by the relative
velocities $\vec v_L- \vec v_i$ and $\vec v_{sl} - \vec v_i$ in
accordance with Galilean invariance. The Lorentz force on the
right-hand side is balanced by the
viscous force $\propto \eta$ and the Magnus force $\propto \eta '$.
The parameter $\eta '$ was known to vary from $\pi \hbar
n_s$ for superclean superconductors \cite{noz} to 0 for dirty
superconductors \cite{kop}. However, $\eta '$ may even exceed $\pi
\hbar n_s$, being equal $\pi \hbar n$, when the Iordanskii force is
essential (see discussion and references in Sec.  X of Ref.
\cite{S}). 

Equations (\ref{elec})-(\ref{vort}) is a closed system of equations averaged
over the vortex-array cell. Equation (\ref{elec}) for electrons does not
differ from the case of the ion lattice at rest. As for Eq. (\ref{ion}) for
ions, I shall show now that the force from vortices in this equation (the
last term on the right-hand side) is required by the momentum conservation
and the third Newton law. 

If there is no external forces on the electron superfluid, the
Helmholtz theorem tells that $\vec v_L = \vec v_{sl}$ \cite{S}. This
means that $\eta=0$ and $\eta '=\pi \hbar n_s$.  Bearing this in
mind, one can present $\vec v_L$ as $\vec v_L = \vec v_{sl}+ (\vec
v_L - \vec v_{sl})$ in the right-hand side of Eq.  (\ref{elec}). Then the
contribution $\propto [(\vec v_L- \vec v_{sl}) \times \vec B]$ is an
external force due to interaction with the normal fluid.
Correspondingly, the same force must appear in the ion (normal fluid)
equation (\ref{ion}), but with the opposite sign. In superfluid
hydrodynamics this force is called {\em mutual friction force}
\cite{S}. But for superconductors it is better to call it {\em
coupling force}, since the force may incorporate not only friction, but
also elastic pinning (see below). The coupling force,
being external for the superfluid component, is internal for the
system ``ions+electrons'' as a whole. Therefore this force doesnot
contribute to the time variation of the total momentum of ions and
electrons
\begin{eqnarray}
\frac{\partial }{\partial t}[m_i n \vec v_i + m_e (n-n_s) \vec v_i +
m_e n_s \vec v_s] \nonumber \\
\approx m_i n {\partial \vec v_i \over \partial
t} + m_e n_s {\partial \vec v_s \over \partial t} =m_in c_t^2 \vec \nabla^2
\vec u_i \nonumber \\ 
+ {1 \over c}[\vec j_l \times \vec B] = \frac{\partial }{\partial z}
\left(m_inc_t^2 \frac{\partial \vec u_i }{\partial z}+ \frac{B\vec h}{4\pi} +
C_{44}^* \frac{\partial \vec u_L }{\partial z}\right) ~. 
           \label{total} \end{eqnarray}
Here $\vec j_l =en_s(\vec v_{sl} -
\vec v_i)$ is the {\em local} current on the vortex line, in contrast
to the {\em average} current $\vec j =en_s(\vec v_s -
\vec v_i)$ in the Maxwell equations (\ref{max}). Equation
(\ref{total}) demonstrates the momentum conservation law for the case
when the wave propagates along vortices (the axis $z$): the time
variation of the momentum is determined by the divergence of the
momentum flux which consists of (i) the stress tensor of the ion
lattice, (ii) the stress tensor of the magnetic field linearized with
respect to the a.c. field $\vec h$, and (iii) the vortex-lattice
stress tensor given by the tilt-modulus $C_{44}^*$. In the
long-wavelength limit $\lambda k \rightarrow 0$ (see below) the
superfluid component $\propto \partial \vec v_s/\partial t$ of the
momentum variation may be neglected, and Eq. (\ref{total}) has a
simpler form:
\begin{equation}
m_i n\left( \partial^2 \vec u_i/\partial
t -  c_t^2 \vec \nabla^2 \vec u_i \right) 
= {1 \over c}[\vec j_l \times \vec B]  ~. 
           \label{total2} \end{equation}
This equation was used in the analysis of Shapira and Neuringer
\cite{SN}, but they neglected difference between the local and 
average electric current which is responsible for the diamagnetism of
the mixed state.

Replacing the coupling force in the equation of ion motion Eq.
(\ref{ion}) with the help of Eq. (\ref{vort}) one obtains:
\begin{eqnarray}
\frac{\partial^2\vec u_i}{\partial t^2} - c_t^2\vec \nabla^2\vec u_i=
-{e \over m_i} {n_s \over n}\left\{\vec E +{1 \over c}[\vec v_i
\times \vec B] \right. \nonumber \\ \left.
- {B \over c}\tilde \eta (\vec v_L-\vec v_i) -{\alpha_M-1 \over c}
[(\vec v_L-\vec v_i) \times B] \right\}~,
                  \label{ion.1} \end{eqnarray}
where $\tilde \eta=\eta /\pi \hbar n_s$ and $\alpha_M = \eta
'/\pi\hbar n_s$ are the dimensionless amplitudes of the viscous and
the Magnus force respectively.  The coupling force has a
component $\propto (\alpha_M -1)$ transverse to the ion velocity
which is maximal in a dirty superconductor when the Magnus force vanishes
($\alpha_M=0$).  Indeed, suppression of the Magnus force in the dirty
superconductor means that impurities produce a force which cancels
the Magnus force on vortices, i.e. on superfluid electrons.  Then in
accordance with the third Newton law one must expect that the force
of opposite sign acts from vortices on impurities, i.e., on crystal
ions. This explains the force $\propto (\alpha_M -1)$ which may be
called the anti-Magnus force.

Later on we assume that the transverse ultrasound plane wave $\propto
\exp(ikz -i\omega t)$ propagates along the 
vortices. From Eqs. (\ref{bv})-(\ref{lon}) one can derive relations
connecting the electric field and the local superfluid velocity with
the ion and the vortex velocities and displacements: 
\begin{equation}
\vec E = - {m_e \over e} \omega^2 \vec u_i - \frac{1}{1+\lambda^2 k^2}
{1 \over c}[\vec v_L \times \vec
B]~,
    \label{E}      \end{equation}
\begin{equation}
\vec v_{sl}= \frac{\vec v_i}{1+\lambda^2 k^2}-\frac{c}{en_sB}
C_{44}(k) k^2 [\hat z \times \vec u_L]~.
             \label{vel} \end{equation}
 where now
\begin{equation}
C_{44}(k)={B^2 \over 4\pi}{1 \over 1+\lambda^2 k^2}+  C_{44}^* 
      \label{c44} \end{equation}
is the $k$-dependent tilt-modulus related to the total energy of
deformation \cite{B}. The contribution $\propto \omega^2$ to the
electric field $\vec E$ will be neglected later on, since it is not connected
with vortices. It yields a small correction to the sound velocity of
the relative order $m_e n_s/m_i n$ which is present even in the
Meissner state. 

Finally, with the
help of Eqs. (\ref{vort}), (\ref{E}), and (\ref{vel}), one can rewrite
the equation of the ion motion, Eq. (\ref{ion}), in the terms of the
ion velocity and displacement only:
\begin{equation}
-\left(\omega^2 - c_t^2 k^2\right)m_in\vec u_i=
  \frac{en_sB}{c} \left\{ f_\parallel \vec v_i
+ f_\perp [\hat z \times \vec v_i] \right\}~.
        \label{ion2} \end{equation}
The longitudinal and the transverse forces on ions from vortices are given
by the dimensionless force parameters:
\begin{eqnarray}
f_\parallel =\frac{Dk^2}{i\omega} -\frac{1}{(\tilde \eta
-Dk^2/i\omega)^2 +\alpha_M^2} 
\left\{\left(\tilde \eta
-\frac{Dk^2}{i\omega}\right)
\right. \nonumber \\ \left. \times
\left[\left( {\lambda^2 k^2 \over 1+\lambda^2 k^2}
\right)^2- \left(\frac{Dk^2}{i\omega}  \right)^2\right] 
+2\alpha_M {\lambda^2 k^2 \over 1+\lambda^2 k^2}\right\}~,
         \label{fpar} \end{eqnarray}
\begin{eqnarray}
f_\perp ={2\lambda^2 k^2 \over 1+\lambda^2 k^2}
+\frac{1}{(\tilde \eta -Dk^2/i\omega)^2 +\alpha_M^2} 
\nonumber \\ \times
\left\{2\left(\tilde \eta -\frac{Dk^2}{i\omega}\right)
{\lambda^2 k^2 \over 1+\lambda^2
k^2} \frac{Dk^2}{i\omega} \right. \nonumber \\ \left. +
\alpha_M\left[\left(\frac{Dk^2}{i\omega}\right)^2-\left( {\lambda^2
k^2 \over 1+\lambda^2 k^2}\right)^2\right]
\right\}~.
         \label{fper} \end{eqnarray}
Here $D=cC_{44}/en_sB$.

The previous experimental and theoretical investigations \cite{P,SN,DB}
dealt with the long-wavelength case  $\lambda k \rightarrow 0$. One may call
it the electrodynamic limit since in this case all forces from vortices on
ions can be expressed in terms of the electrodynamic parameters: the Ohmic
and the Hall conductivities, $\sigma_O=\eta c^2/\Phi_0B$, $\sigma_H=\eta '
c^2/\Phi_0B$, related to the viscous and the Magnus force respectively, and
the magnetic permeability
\begin{equation}
\mu=\frac{B^2/4\pi}{B^2/4\pi +C_{44}^*} = \frac{B^2}{4\pi C_{44}(0)}~,
     \label{mu} \end{equation}
which describes the diamagnetism due to
circular currents over the vortex-lattice cell \cite{S2}. Here $C_{44}(0)$ is
the tilt-modulus in the limit $k \rightarrow 0$ (the Labusch
modulus). One can check that Eq. (\ref{mu}) yields the differential
magnetic permeability $\mu =B/H$ for the case when the variation of
the magnetic field is normal to the vortices. Here $H=(1/4\pi)
\partial F(B)/\partial B$ is the thermodynamic magnetic field along
the equilibrium magnetization curve. In the terms of electrodynamics
the equation of vortex motion (\ref{vort}) is simply the Ohm law
\begin{equation}
\vec j_l= \sigma_O \vec E_i -\sigma_H[\hat z \times \vec E_i]
  \label{ohm}~,       \end{equation}
which connects the local current $\vec j_l = \vec j /\mu$ and the electric
field $\vec E_i = \vec E +(1/c)[\vec v_i \times \vec B]$ in the coordinate
frame moving with the ion velocity. Then Eq. (\ref{ion2}) may be derived
from  Eq. (\ref{total2}) together with the Ohm law Eq. (\ref{ohm}) and  the
relation $\vec E =(4\pi i\omega/c^2k^2) \vec j=(4\pi i\omega\mu/c^2k^2) \vec
j_l $  which follows from the Maxwell equations (\ref{max}). This yields the
following values of the force parameters:
\begin{eqnarray}
f_\parallel =\frac{B}{en_sc}\left( \frac{c^2k^2}{4\pi i\omega\mu}\right)^2
\left[ \frac{4\pi i\omega\mu}{c^2k^2} \right. \nonumber \\ \left.
+\frac{\sigma_O-c^2k^2/4\pi
i\omega \mu}{\left(\sigma_O-c^2k^2/4\pi i\omega \mu
\right)^2+\sigma_H^2} \right]~, 
               \label{fpar0} \end{eqnarray}
\begin{eqnarray}
f_\perp =\frac{B}{en_sc}\left( \frac{c^2k^2}{4\pi i\omega\mu}\right)^2
\frac{\sigma_H}{\left(\sigma_O-c^2k^2/4\pi i\omega \mu
\right)^2+\sigma_H^2} ~.
         \label{fper0} \end{eqnarray}
One can easily check that Eqs. (\ref{fpar0}) and (\ref{fper0}) coincide with
Eqs. (\ref{fpar}) and (\ref{fper}) if $\lambda k \rightarrow 0$. In the
electrodynamic limit our theory is valid in a wider interval of the magnetic
fields than in the general case of arbitrary $\lambda k$. Since we use the
equations averaged over the vortex-array cell, they hold until the
wavelength $2\pi/k$ exceeds the intervortex distance $a \sim
\sqrt{\Phi_0/B}$. But because of $\lambda \ll 1/k$, this condition may be
satisfied   even  if $a \gg \lambda$, i.e., rather close to the lower
critical field $H_{c1}$ where the diamagnetism is important, i.e., $\mu$
essentially less then unity. On the other hand, in the electrodynamic
approach our assumption in the beginning of the paper that the current of
normal electrons is negligible is not necessary: one can use the
conductivity tensor taking into account this current. Then the theory is
valid even  close to $H_{c2}$.

For the sake of simplicity we did not include the elastic pinning
force into our analysis explicitly, but it is easy to do simply by
replacing the viscous coefficient $\eta$ by $\eta -\alpha_P/i\omega$
in all equations. Here $\alpha_P $ is the bulk pinning coefficient
which may, in principle, depend on the frequency as assumed by
Dominguez {\it et al.} \cite{DB}. Thus our analysis holds for any
regime of vortex motion, either the non-activated flux flow,
or the thermally assisted flux flow (TAFF) with flux jumps over
pinning barriers. But different regimes of the flux flow correspond
to different expressions for the conductivities.

One can obtain the results of Refs. \cite{P,SN} from Eqs.
(\ref{fpar0}) and (\ref{fper0}) neglecting the Magnus force and 
the finite diamagnetism of the mixed state ($\sigma_H \sim
\alpha_M=0$, $\mu=1$). Note that the magnetic
permeability $\mu$ of Shapira and Neuringer \cite{SN} relates to the
atomic magnetism, whereas the latter was ignored in the present
work and our $\mu$ is due to the circular currents in the
vortex-lattice cell and the difference between the local and average
currents as a result of them. However, finally the role of $\mu$ in
the equations is similar in both cases, as one might expect for an
electrodynamic theory.

Our analysis seriously differs from that of Dominguez {\em et al.}
\cite{DB}: (i) They missed to take into account the anti-Magnus force in the
equation of ion motion. This caused violation of the momentum conservation
law and the wrong prediction for the Faraday effect in the dirty
superconductors in which the anti-Magnus force is especially important. (ii)
The equation of vortex motion used in Ref. \cite{DB} contained the
laboratory vortex velocity and thereby was not Galilean-invariant.  This
means an assumption that the laboratory frame is preferential.  But for our
problem the only preferential coordinate system is that moves with the ion
velocity.

Experiment and theory \cite{P} have shown that  in a Bi superconductor there
is a crossover between two temperature regions: the low-temperature region
of high Ohmic conductivity due to high activation barriers in the TAFF model
where $\omega \tilde \eta/Dk^2 =c_t \tilde \eta/Dk=
4\pi\omega\mu\sigma_O/c^2k^2 > 1$, and high-temperatures region of low
conductivity where $4\pi\omega\mu\sigma_O/c^2k^2 < 1$.  The predictions of
Dominguez {\em et al.} \cite{DB} for the Faraday effect must be revised both
in the low-temerature and the high-temperature regions. At low temperatures
they obtained that the Faraday rotation (the angle of polarization rotation
per unit length) is $d\theta/dz = \sigma_H B^2/2m_inc_t c^2$, whereas Eq.
(\ref{fper0}) yields the Faraday rotation $d\theta/dz=en_sB
\mbox{Re}f_\perp/2m_i nc_t c$ which is by the factor
$(c^2k^2/4\pi\omega\mu\sigma_O)^2$ smaller. This factor is of order unity at
the crossover temperature $T \sim 60$ K, but at low temperatures the Ohmic
conductivity increases proportionally to $ \exp (U/T)$ where the activation
barrier is about 500 K for typical magnetic fields \cite{P}. Thus Dominguez
{et al.} overestimated the Faraday effect at low temperatures  by many
orders of magnitude. At higher temperatures  Dominguez {et al.} obtained the
Faraday rotation $d\theta/dz = eB/2m_i c_tc$ which did not depend on the
Hall conductivity. This means that the Faraday effect is possible in a dirty
superconductor without the Magnus force even in the limit  $\lambda k
\rightarrow 0$ in disagreement with Refs. \cite{P,SN}.  Dominguez {et al.}
explained it by the effect of the electromagnetic force $\propto \left(\vec
E+ {1 \over c}[\vec v_i \times \vec B] \right)$ [see Eq. (\ref{E}) for $\vec
E$] in Eq. (\ref{ion.1}), which was neglected in  Ref. \cite{P}. However, as
explained after Eq. (\ref{ion.1}), this equation should include also the
anti-Magnus force which cancels the electrodynamic force  in the limit
$\lambda k \rightarrow 0$ restoring the result of Refs. \cite{P}. According
to Eq. (\ref{fper}) the Faraday rotation without the Magnus force ($\alpha_M
\sim \sigma_H=0$) is given by  
\begin{equation} 
\frac{d\theta}{dz}=\frac{en_s B}{2m_i n c_t c} \mbox{Re}  f_\perp =
\frac{en_s B}{m_i n c_t c}\, \frac{\lambda ^2k^2}{1+\lambda ^2k^2}\,
\frac{1}{1 +  (c^2k^2/4\pi\omega\mu\sigma_O)^2} ~.
         \label{short} \end{equation}
For the ultrasound frequencies 3 MHz from Ref. \cite{P} this expression
yields at high temperatures $>60$ K the Faraday rotation at least by the
factor $10^{-5}$ smaller than predicted by  Dominguez {\em et al.}
\cite{DB}. So the Faraday effect in dirty superconductors  is possible only
at $\lambda k \geq 1$. However, the frequency of ultrasound must be very
high in order to achieve this case (about a few GHz). For the frequencies
used in the ultrasound experiments now, the best conditions for observation
of the Faraday effect are clean superconductors with large Hall angle at
high temperatures. Then the Faraday rotation can be strongly amplified close
to the resonances with vortex modes. The resonances correspond to zeroes of
the denominator in Eqs. (\ref{fpar0}) and (\ref{fper0}). Weakly damped
vortex modes exist in superclean high-T$_c$ superconductors according to
recent experimental and theoretical investigations \cite{KS,O}.  Thus the
ultrasound experiment is able to reveal these vortex modes as has been
already discussed in Ref. \cite{BI}.

In summary, the theory of interaction between the ultrasound and the
vortices in the type-II superconductors has been suggested. The first time
the theory takes into account the Magnus force on vortices, the anti-Magnus
force on ions, and the diamagnetism of the mixed state self-consistently. 
This results in a serious revision of previous predictions concerning the
acoustic Faraday effect. Despite this revision, our analysis confirms that
possible observation of the acoustic Faraday effect is expected to provide a
valuable information on vortex dynamics in type-II superconductors.

I thank Gianni Blatter, Vadim Geshkenbein, Nikolai Kopnin, Anne van Otterlo,
and Konstantin Traito for useful discussions. I appreciate comments on the
paper by Lev Bulaevskii. The present work has been partially supported by
the Soros International Foundation and by the Russian Foundation for
Fundamental Investigations (grant No. 94-02-5951).

\end{document}